# MOIST: A Scalable and Parallel Moving Object Indexer with School Tracking

Junchen Jiang, Hongji Bao, Edward Y. Chang and Yuqian Li
Google Research
{jasonjiang, hongjibao, edchang, liyuqian}@google.com

## ABSTRACT

Location-Based Service (LBS) is rapidly becoming the next ubiquitous technology for a wide range of mobile applications. To support applications that demand nearest-neighbor and history queries, an LBS spatial indexer must be able to efficiently update, query, archive and mine location records, which can be in contention with each other. In this work, we propose MOIST, whose baseline is a recursive spatial partitioning indexer built upon BigTable. To reduce update and query contention, MOIST groups nearby objects of similar trajectory into the same school, and keeps track of only the history of school leaders. This dynamic clustering scheme can eliminate redundant updates and hence reduce update latency. To improve history query processing, MOIST keeps some history data in memory, while it flushes aged data onto parallel disks in a locality-preserving way. Through experimental studies, we show that MOIST can support highly efficient nearest-neighbor and history queries and can scale well with an increasing number of users and update frequency.

## 1. INTRODUCTION

The number of "smart" wireless devices such as mobile phones and iPad-like devices have been rapidly growing. Being able to keep track of the locations of moving devices can enhance a number of applications. Location-Based Service (LBS) is rapidly becoming the next ubiquitous technology for a wide range of mobile applications, such as location positioning, location navigation, location-aware search, social networks, and advertizing, just to name a few. Besides the traditional "where am I" queries, LBS must support nearest-neighbor (NN) queries and history queries. Some examples of NN query are: when a store wants to find nearby customers to offer coupons, a user wants to find nearby friends, or a taxi driver wants to locate nearby clientele. History queries can find popular routes of pedestrians, and locate points of interest in a city. A delay in providing accurate NNs can degrade user experience or cost business opportunities. Both NN and history queries on moving objects bring technical challenges to LBS design.

Performing NN queries on static objects can be supported by implementing a traditional spatial indexer (see Section 2 for representative work). Indeed, Google Maps has been using a spatial indexer called *S2Cell* to index geographically nearby objects. However, when objects can move, and at the same time users are interested in their history as well as current locations, an indexer must satisfy performance requirements under new constraints. The first requirement (constraint) is that the *update latency* must be short. (Here, we define *update latency* to be the time from when an update is requested by a mobile device to the time when the location is available on the indexer to be queried.) The issue at hand is that huge numbers of update queries usually results in long update latency, which in turn can cost user convenience. For instance, a one-minute update latency in NYC can be translated into a distance of a couple of city blocks traveled by a bus or taxi. The second performance requirement is related to historical queries. Not only must history be made available for queries, it should also be made available quickly. The indexer must thus cache some history data in memory and, at the same time, flush out aged data onto disks to vacate memory space for new location updates.

To reduce update latency and to support efficient history queries, we propose MOIST (Moving Object Index with School Tracking). On top of Google BigTable and spatial indexer, MOIST performs object schooling (explained shortly) to eliminate redundant updates. Aged data is treated to preserve its on-disk locality through a parallel ping-pong scheme. More specifically, MOIST consists of four key schemes:

1. *Key-value model* is the baseline of MOIST. MOIST uses BigTable to store the three records, namely the Location Table, Spatial Index Table and Affiliation Table.

2. *Spatial indexer*. MOIST uses the Google S2Cell indexer to hierarchically decompose a space into cells of different resolutions. The cells have a space-filling curve structure that makes them efficient for spatial indexing. An arbitrary region can be approximated by a collection of cells. When performing an NN query, MOIST enhances S2Cell to adapt to cells' population density to look up NNs in cells of different resolutions.

3. *Object school*. On top of the key-value pairs, MOIST clusters data into object schools (OSes) consisting of objects that are close in geographic proximity and similar in velocity. OSes are pervasive among urban-area moving objects such as passengers on cars on freeways. OSes reduce redundant updates and hence cuts down update latency. (Section 2 presents an example to illustrate the key differences between OSes and traditional object clustering approaches.)

4. *Aged data archiving*. MOIST employs PPP, a parallel ping-pong scheme, to flush history data onto disks. When doing so, PPP attempts to preserve data locality so as to support efficient on-disk location-based and object-based history queries.

Our experimental results show that MOIST achieves much higher Queries Per Second (QPS) than the best record of previous approaches by one, or in some cases, even two orders of magnitude. Take update as an example. With only one server accessing BigTable and no object school (clustering), MOIST achieves $8,000+$ updates per second with one million moving objects (2x better than $3,000+$ QPS of $B^x$-tree [15] even when only input/output (IO) time is counted). With 10 servers and object schools, MOIST achieves update QPS of 60k and about 80% of the updates generated in a road-network map are shed by object schools, showing a nearly 80x speedup over $B^x$-tree.

In summary, this paper makes two significant contributions to the design of a location-based data service:

1. *Traffic shedding*. MOIST reduces update latency through exploring moving-pattern correlations between moving objects, and thus eliminates redundant updates and storage.

2. *History data archiving*. MOIST employs PPP to ensure efficiency on both aged-data archiving and history queries.

The rest of the paper is organized as follows: In Section 2, we survey related work. Section 3 presents MOIST in detail, including its baseline. Section 4 reports experimental results. Section 5 briefly presents a deployed application that uses the techniques introduced in this paper. Finally, we offer our concluding remarks in Section 6.

## 2. RELATED WORK

We address three problems in this work: spatial indexing, traffic shedding, and object clustering. We discuss related work accordingly.

### 2.1 Spatial Indexers

Moving object indexing traditionally employs R-tree or B-tree structures. An R-tree employs a maximal bounding rectangle (MBR) to bound location of the objects in a subtree. A TPR-tree, [23] and later, TPR*-tree [24] are proposed to support temporal services (predictive or historical queries). In a query-intensive scenario, an R-tree structure and its derivations are more likely to outperform other B-tree based methods, as the MBRs employed by R-trees are of the same order of the number of mobile devices and a large part of these are pruned for each query. However, in update-intensive cases, R-trees tend to spend more time on maintaining target structures than B-trees. Two representative methods to implement temporal service are $B^x$-trees [15] and $B^{dual}$-trees [27]. $B^x$-trees do not use any MBR and index objects by using space-filling curves [20] to serialize a 2-D space into a 1-D key space. $B^{dual}$-trees partition both spatial and velocity spaces and thus key objects in a four dimensional space. In [19], a $B^x$-tree forest is created to enable more efficient historical and predictive queries. Like B-tree based methods, MOIST uses space-filling curves as its baseline spatial index. However, MOIST leverages fine-grain tuning on scan size using a method introduced in Section 3.4, and it also requires no disk- or memory-based optimizations as in [2] or [7], because the BigTable infrastructure of MOIST uses a finer granularity when fetching each cell and support for range scan [5].

### 2.2 Traffic Shedding

Many previous works throttle workload to achieve higher performance by shedding updates on the records of a single user. QU-trees [25] shed index updates by controlling the underlying index's filtering quality. An alternative could be to shed updates using a Kalman Filter [14]. MobiQual [11] has the same targets while taking skew in QoS of different updates and queries into account. In the recently proposed STSR [8], the same workload shedding is performed in a decentralized protocol run by each user, shedding unnecessary updates before they are sent to the server. Recent work [21] scales location updates with so-called active shedding by introducing query encounter points to make the traffic reduction QoS-aware. *Safe region* is used as a common base solution in many proposals to carry traffic shedding [3] [13] [22]. In contrast to these approaches, MOIST sheds updates by exploiting relationships between users, rather than making use of the data of just a single user (like in [25] [11] [8]).

### 2.3 Object Clustering

Prior works on clustering can be divided into two approaches: *static* and *dynamic*.

#### 2.3.1 Static Clustering

A number of common patterns can be defined to be represented by cluster prototypes. A moving segment is then represented by one of these prototypes, and a long-term moving trajectory consisting of several moving segments is represented by a sequence of prototypes. The benefit of this kind of static clustering approach is its efficiency in dealing with pattern changes. Some shortcomings however, are that the pre-defined prototypes may not cover all possible patterns, and an approximation of the trajectory causes information loss. Works of *coreset* proposed in [12] and [9] are representative static approaches. Mapping a moving object with an uncertain behavior model is studied in [28].

#### 2.3.2 Dynamic Clustering

Dynamic clustering finds similar patterns in real time, and represents a cluster by a virtual center moving in linear model and a radius [16], or by a bounding rectangle [18]. Each object in a cluster shares the same moving pattern as that cluster. A cluster's moving pattern is influenced by each object's updates, based on which the moving pattern is adjusted or the object departs the cluster. Our proposed *object school* is also a dynamic clustering scheme that differs from [16] and [18] in that we keep track of only one object's location changes, and shed updates of all others. Therefore, the number of updates MOIST needs to handle is *independent* of the cluster size. Imagine the most optimistic case, where all subway passengers can be tracked by just one single lead passenger. All clustering algorithms face the challenge of overhead during a merge and split operation caused by pattern changes. We will discuss this in Section 3.3.

### 2.4 Illustrative Example

Figure 1 shows how our object-school method differs from the afore-mentioned traditional clustering methods. The figure presents two objects, *A* and *B*, moving along a road (depicted between two curved lines) in a similar trajectory. Object *A* makes four turns, whereas object *B* makes five. Suppose a static clustering scheme may use four predetermined moving patterns (prototypes) to describe *A* and *B* (in Figure 1(a)). Both *A* and *B* must be reclassified into a predefined prototype when a turn is made. Both their locations must be updated in their spatial indexer. Next, a dynamic clustering scheme (in Figure 1(b)) like the ones proposed in [16] and [18] may decide to cluster *A* and *B* (and some other objects) into one cluster, and create a virtual center to represent them. This re-clustering is performed at each major turn, and therefore, we can see in Figure 1(b) that three large arrows, representing three virtual centers, are generated along the path. This reclustering and virtual center computation involves all objects and can be both computational intensive (at least $O(n \log n)$, where *n* is the number of objects) and IO intensive (reading each object's moving pattern). Our OS scheme finds a *leader* (*A* is chosen to be the leader in Figure 1(c)) to represent a school (cluster), and location updates of a school involve only one object on the indexer. This school clustering eliminates a tremendous amount of redundant updates on the indexer. Furthermore, our re-clustering algorithm can enjoy an $O(\log n)$ complexity since it uses the existing leaders as candidate prototypes to align objects.

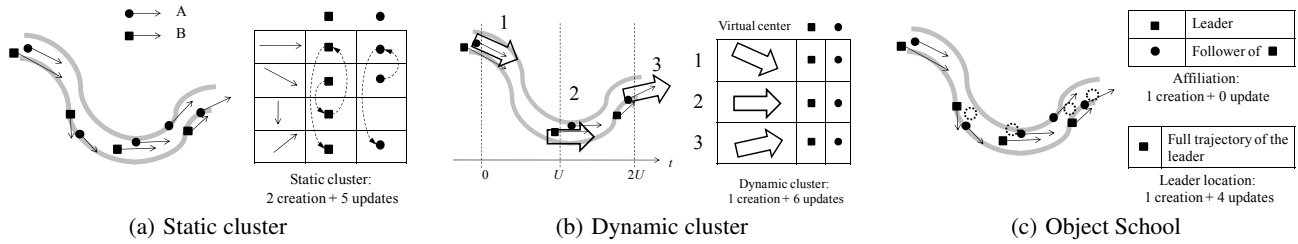

(a) Static cluster  (b) Dynamic cluster  (c) Object School

**Figure 1: Comparison of static cluster, dynamic cluster and object school**

## 3. MOIST

As presented in Section 1, MOIST aims to reduce update latency and support efficient history queries. MOIST consists of four key components: *key-value-based data structures*, *spatial indexers*, *object schooling*, and *aged-data archiving*.

### 3.1 Data Structures

Before presenting detailed schemas of the two key tables (i.e., the *Location Table* and *Affiliation Table*), we briefly depict some benefits of using BigTable. Our use of BigTable leads to the essential difference between our storage method and that of previous works (e.g., [24] [7] [15]). First, BigTable is a key-value based distributed storage system that stores and manages values by sorting row keys, and the values can be configured to store in memory or on disk by columns. Second, BigTable supports batch reading for accessing values from a set of contiguous key blocks, and this reading method performs much faster.

The Location Table stores the location and velocity information for each object, so that queries on an individual object can be answered efficiently. To track object schools (OSes), MOIST maintains the Affiliation Table to record the mapping for each cluster and is keyed by the behavior (i.e., the leader's velocity) of that cluster. The Spatial Index Table stores the leaders by their spatial locations which will be detailed in Section 3.2.

#### 3.1.1 Affiliation Table

The Affiliation Table of MOIST is keyed by an Object ID (OID). The goal of the Affiliation Table is to provide fast access – for a leader to locate its followers and for a follower to find its leader.

The design of the Affiliation Table includes two basic operations: first, to check if an object is a leader or a follower; and second, in the case it is a follower, to calculate its estimated location by its displacement from its leader. The Affiliation Table implements these two operations by two column families, namely *L/F* and *Follower Info*.

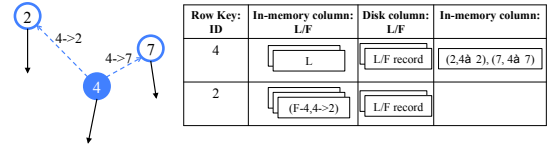

**Figure 2: Schema example of Affiliation Table**

*L/F records*: A leader only has one L/F record indicating that it is a leader, and a timestamp indicating the time the leader is chosen. A follower *j* has multiple L/F records with the same content (F-*i*, *i* → *j*) where *i* is its leader. The L/F record will be renewed each time the follower updates its location. Fresh L/F records are stored in the in-memory column, and after a period of time, aged L/F records will be transferred to disk columns.

*Follower Info*: The Follower Info of a leader *i* is a concatenation of pairs in the form of $(j, i \to j)$, where *j* is one of its followers and $i \to j$ is the vector replacement in location from *i* to *j*. Only leaders have content in the Follower Info column and this content does not change until either any follower leaves the OS or the leader becomes a follower itself.

The Affiliation Table also reduces the workload on the Location Table by pre-processing queries, and only passes them to the Location Table (see the next section) if those queries are irrelevant to school (cluster) information.

#### 3.1.2 Location Table

The Location Table is employed to keep necessary information for each object. The Location Table sets each row's key as the object ID (OID) of a particular object. Each row then stores multiple location records of the corresponding object at different times. In Location Table, each location record includes various information such as location, velocity, etc of the object. The Location Table contains one in-memory column and several disk columns. Each location record is timestamped and stored in such a way that the most up-to-date location records are kept in the in-memory column. After a fixed period of time, all records are then considered aged. The location records in each column are then compressed and transferred to the next disk column (The scheme for aged data transfer is covered in Section 3.5.) Figure 3 depicts an example schema of a Location Table.

#### 3.1.3 Example Tables

Figure 4 shows an example of three simplified tables for 6 objects grouped into 2 OSes. As objects 4 and 6 are leaders, the Location Table and Spatial Index Table only hold records for objects 4 and 6. The Affiliation Table keeps track of all leaders and followers, as well as the displacements from each leader to its followers.

| Row key: ID | Column family: Location signal | | |
|---|---|---|---|
| | In-memory column: Location | Disk column: Location | Disk column: Location |
| Alice | Location record | Location record | Location record |

**Figure 3:** Example schema of a Location Table, where location records include location, velocity, etc. Each record is timestamped. The Location Signal Column Family contains one in-memory column and several disk columns in order to store data aged to different degrees.

**Figure 4:** The content of a Location Table, Spatial Index Table and Affiliation Table for 6 objects grouped into 2 OSes

## 3.2 Spatial Indexer

We first depict a spatial indexer for static objects, and then present our enhancements to index moving objects.

### 3.2.1 Indexer for Static Objects

An example for querying static objects can be that of a mobile user who asks for the locations of nearby bus stops or coffee shops. To query static objects, a server maintains the location of all objects using an index structure called a *Spatial Index Table*. Hereafter, we refer to the specific Google implementation of this as S2Cell which uses Hilbert Curves [20] to form a B-tree indexing scheme. From a bird's eye view, the Spatial Index Table first linearizes a 2-D space into 1-D key space by partitioning the space into a $2^n \times 2^n$ grid and numbering each cell with a key called a *spatial index*. If a static object is located within a given cell, then that object's ID is stored as the value associated with the key of that cell. Then, any query for the static objects on 2-D space can be transformed to a combination of queries on the 1-D key space for which BigTable provides parallelism to read data from multiple ranges.

Mathematically, suppose that the two-dimensional space is $[0,1]^2$ and the spatial index function is $h(\cdot): [0,1]^2 \to [0,1]$. The function $h((x,y))$ can be easily achieved with division and encoding: divide $[0,1]^2$ into four small squares of length $2^{-1}$, encode these four squares by $00, 01, 10, 11$, find the small square which contains location $(x,y)$, write its encoded number as $d_1 \in \{00, 01, 10, 11\}$, then divide $d_1$ into four smaller squares of length $2^{-2}$, encode and find a smaller square $d_2 \in \{00, 01, 10, 11\}$, and so on so forth. After $l$ rounds of this process, one can map each location to a finite sequence $(d_1, d_2, d_3, \ldots, d_l)_2$. Finally, define $0.d_1d_2d_3\ldots d_l$, the binary fraction, as the spatial index value where $l$ is called the *level* of the spatial index:

$$h((x,y)) = (0.d_1d_2d_3\ldots d_l)_2 \in [0,1]$$

A space-filling curve is constructed by linking all cells in a space by the sequential order of their spatial indexes. Hilbert Curves [20] are a representative scheme of space-filling curves, which guarantee locality, a property that cells geographically close together are likely to be given spatial indexes close in value in the key space. We use Hilbert Curves to encode spatial indexes. This is illustrated in the left-hand side of Figure 5. While other encodings such as Z-curves are also applicable to perform space-filling, Hilbert Curves perform slightly better [15].

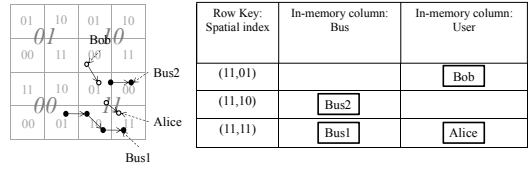

**Figure 5:** Example scheme of a Spatial Index Table. The left side is an example of how space is recursively divided and encoded by spatial indexes. The right-hand side is the corresponding rows in the Spatial Index Table.

In the case of the surface of the Earth as 2-D space (which lies on a positively curved geometry, as opposed to flat Euclidian geometry), the 2-D surface is first partitioned into six square parts, and Hilbert Curves are employed to each part. On each part of these six, mathematical methods are used to normalize Earth's curved surface onto a 2-D plane surface [17]. It is only after this process that objects on the Earth's surface can be indexed.

Given a location coordinate, it is first mapped into one of the six squares described earlier. Then, its spatial index is calculated at the first level (i.e., $d_1$) and then $d_i$ ($i = 2, \ldots, l$), recursively. Finally, they are concatenated to obtain the spatial index. The reverse direction, i.e., calculating a location coordinate from a spatial index, is similar. In this way, a location update can be instantly mapped to the row that is keyed by its corresponding spatial index; this allows the distance between two spatial cells to be easily calculated. An example of a Spatial Index Table is shown in the right side of Figure 5.

### 3.2.2 Challenges of Moving Objects

Querying over moving objects is much more challenging than querying over static objects. The reason is simple, when objects move, location updates can be in contention with location queries. In addition, when a server has to handle a large number of updates, and when the volume of those updates increases with the population of objects, or the update frequency of each object is escalated, the number of updates will eventually overwhelm the capacity of BigTable. This scalability issue can cause amplified update latencies as we depicted in Section 1. MOIST addresses the update latency problem by employing object schooling, which we present next in detail.

## 3.3 School Clustering

MOIST clusters nearby objects with similar moving patterns into one school. This mapping is based on the observation, called *Object Schooling*, that nearby objects often move with similar velocities, and more importantly, with similar trajectories. Take objects on a subway for example, the static clustering approach may require re-clustering whenever the subway turns. The dynamic approach may not consider the "nearby" factor causing a cluster to consist of objects which are far apart (though with the same moving pattern). In contrast, MOIST calls a group of nearby objects moving in concert an *object school (OS)*. MOIST keeps track of an OS by tracking only the *lead* object, and records the distance between locations of the other *follower* objects and the leader for precise location query on each object.

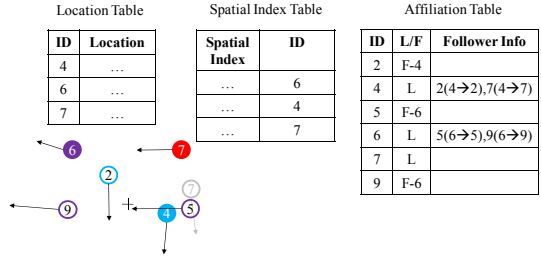

Figure 6: Example of updating

OSes are maintained and renewed when a location update arrives. An object is considered to have departed its OS when its displacement to its leader exceeds a certain threshold. When an object leaves an OS, it becomes a leader of a new OS. We periodically merge OSes using a quick cluster method. Next, we discuss how location updates may affect three data structures of MOIST, namely the Location Table, the Spatial Index Table and the Affiliation Table.

### 3.3.1 Location Updates

An update message consists of an object's ID, location and velocity. Upon receiving an update request, MOIST updates the Location, Spatial Index and Affiliation tables by Algorithm 1, which has three basic branches: one for a leader, one for a follower that remains within a school, and one for a follower that departs an OS.

The location update procedure begins by checking if the update is a leader or a follower–this is accomplished by looking up the L/F record of its *ID* in the Affiliation Table (line 1). If the object is a leader, the update will be done by changing both the Location Table and the Spatial Index Table (line 1,1). Line 1 ensures that each ID has one record in the Spatial Index Table. In the case that the object is a follower, MOIST first decides whether the update can be shed by analyzing the encoded location. If the estimated location (to be shown soon) of the follower is close enough to the real location (i.e., within a threshold $\varepsilon$), the update is shed; otherwise, the object is considered as having left the school. In this case, we transform *ID* into a leader and update its records by writing its location information in the Location Table and insert its ID to the corresponding spatial index in the Spatial Index Table (line 1 to 1).

---

**Algorithm 1** MOIST Update Procedure

**Require:** 4-tuple (ID, $\overrightarrow{Loc}, \overrightarrow{V}, t$), location $\overrightarrow{Loc}$, velocity $\overrightarrow{V}$
1: **if** ID is a leader **then**
2:    Location Table: add ($\overrightarrow{Loc}, \overrightarrow{V}$) to row ID with timestamp $t$
3:    Spatial Index Table: delete ID in previous spatial index and add it into new spatial index
4: **else**
5:    Find ID's leader $i$ in L/F column in Affiliation Table
6:    Calculate ID's estimated location $\overrightarrow{Eloc}$ in $t$
7:    **if** $Distance(\overrightarrow{Eloc}, loc) \leq \varepsilon$ **then**
8:       Shed update and do nothing
9:    **else**
10:      Delete ID's record in $i$'s Follower Info
11:      Affiliation Table: Label ID as leader by adding new L/F entry with timestamp $t$
12:      Location Table: add ($\overrightarrow{Loc}, \overrightarrow{V}$) to row ID with timestamp $t$
13:      Spatial Index Table: add ID into new spatial index
14:    **end if**
15: **end if**

---

The *calculation of estimated location* involve four steps: (i) find the object's leader $i$ (line 1), (ii) get $i$'s latest location record (including $\overrightarrow{Loc}$ and $\overrightarrow{V}$) from the Location Table, (iii) calculate $i$'s location $\overrightarrow{Loc'}$ at time $t$, and (iv) get $\overrightarrow{Eloc} = \overrightarrow{Loc'} + (i \rightarrow ID)$, where $i \rightarrow ID$ is the displacement from leader $i$ to follower *ID*. Thus, an OS consists of a leader $L$ and followers $F$, and is formally defined as

$$\{F | Distance(\overrightarrow{Loc}, \overrightarrow{ELoc}) < \varepsilon\}$$

Figure 6 shows an example where object 7 is too far away from its estimated location. When receiving an update from object 7, MOIST recognizes it is a follower and calculates its estimated location, denoted by a gray circle. Since the updated location digresses by a margin exceeding our threshold, we let object 7 be the leader of a new OS and delete its record in its former leader 4.

### 3.3.2 Clustering

Clustering is executed a periodically over a region of a given level of spatial cell, called a *clustering cell*. Since a clustering cell is several levels higher than the spatial cell used in the Spatial Index Table, spatial cells of a clustering cell will have consecutive spatial indexes. As a result, one can retrieve all leaders in a clustering cell quickly from the Spatial Index Table by leveraging batch reading of BigTable. It then groups leaders having similar velocities into one OS with only a small overhead on time and computation. Two velocities are considered similar if the value of their vector difference is less than a threshold.

Within each clustering cell, we require the time for clustering to be $O(n)$ where $n$ is the number of OSes in the cell. We define the velocity space as a 2-D space, where any velocity can be projected by fixing its start point on the center of the space. We use $\Delta_m$ to denote the maximum velocity deviation within an OS. We first partition the velocity space into identical hexagons as shown in Figure 7 (on the right-hand side), which guarantees that the maximum distance between two internal points is less than $\Delta_m$. To cluster all OSes in the cell, each leader is first mapped to the corresponding hexagon partition in $O(1)$ time. After we map all the leaders, those within one hexagon will be merged into one OS. To merge one leader $j$ and its followers into another leader $i$ incurs three operations: (i) to transfer all the Follower Info of $j$ into $i$, (ii) to change the L/F entry for all of $j$'s followers, and (iii) to delete leader $j$ from the Spatial Index Table.

Figure 7 shows an example of clustering where three OSes (the same as those OSes in Figure 6) are grouped into two OSes. First, we retrieve all leaders from the Spatial Index Table: 4, 6, and 7. Then, cluster the leaders by their velocity. In Figure 7, we find that 6 and 7 should be merged into one cluster and 4 in another. Object 7 is merged into leader 6.

To reduce the impact of clustering the entire map, we sequentially cluster each clustering cell, so at any given time only a small number of clustering cells are being processed. To explain the advantage of this scheme, we consider another two clustering methods: static and dynamic clustering as introduced in Section 2.3. Static clustering [12] [9] finds for an object a new leader once it departs from a school. Prior dynamic clustering methods use trajectory prediction (e.g., linear movement [16] or micro-cluster [18]) to avoid clustering and do reclustering on-demand. While these methods all incur too many read/write operations with BigTable if objects leave schools or change moving patterns frequently, our method aggregates the operations into one clustering, and reads and writes BigTable in batch, reducing networking overhead greatly. Our advantage over static and dynamic clustering would be even more clear in extreme cases where moving patterns of objects change frequently as discussed next.

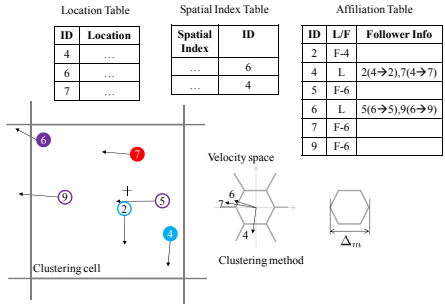

Figure 7: Example of clustering

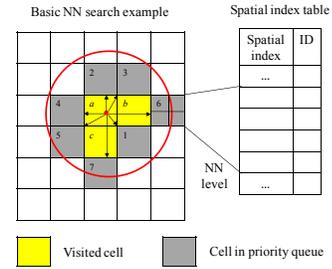

Figure 8: The example of nearest neighbor search algorithm

### 3.3.3 Dealing with Extreme Cases

Since an OS is maintained by comparing the velocity of each follower with that of the single leader, if the leader is representative enough, it costs only light overhead to shed a large amount of updates. However, an extreme case could be that at some moment all members in an OS change their behavior simultaneously so that no object can be representative enough to become the leader. E.g., when a bus stops, many passengers of one OS who were moving together at the same velocity, leave towards different directions at almost the same time.

To handle these extreme cases, static clustering finds, as soon as possible, new representative patterns (leaders) from the objects, i.e., performs re-clustering immediately. However, if the objects are changing velocity (e.g., some are transferring from one metro to another, some are leaving the station), the chosen pattern (leaders) cannot be representative for enough time, leading to more re-clustering. In addition, to detect such phenomenon from a large amount of updates will be costly. Dynamic clustering predicts the moving patterns of objects and performs re-clustering on demand. It suffers from the same problem that re-clustering might be intensive as each object's moving pattern is unstable.

In contrast, our approach (in Section 3.3.2) is to perform lazy clustering, i.e., to do re-clustering periodically and not to detect such phenomenon. Our method prevents unstable clustering results in two ways. First, we take into account the geographical proximity when judging if a follower leaves the OS. If a follower is near the leader, it is still within the OS even if it changes the moving pattern radically (e.g., most passengers just leaving a metro will still be in geographical proximity for a while), avoiding unstable clustering resulting from objects' changing moving patterns. Second, we do re-clustering periodically, saving the cost of detecting the extreme cases. In fact, the interval between two re-clusterings becomes a tradeoff which is studied in the section of experiments (4.2.2). Finally, although our clustering method takes the risk of having an increased number of OSes since the unclustered objects will become leaders and form new OSes, we argue that it is unlikely that such extreme cases occur at the same time and at the same place, i.e., re-clustering behavior will be "local" not global. For instance, only a fraction of buses stop at a given time and only a fraction of passengers get off/on a bus/subway at a given time.

## 3.4 Nearest Neighbor Search

Nearest neighbor query enables clients to retrieve the $k$ nearest objects around a given location $loc$. It has four steps: (i) estimate the number of leaders to be retrieved, (ii) find a certain number of the nearest leaders around $loc$ from the Spatial Index Table, (iii) fetch the followers of these leaders, and (iv) calculate locations (of the leaders/followers) and return the $k$ nearest ones. In this part, for better demonstration, we'll first present the algorithm for finding a certain number of the nearest leaders, then introduce how to estimate the number of leaders we need to retrieve, and at last, describe the way to extend this algorithm to predictive nearest neighbor search.

### 3.4.1 Nearest Leader Search Algorithm

We introduce the algorithm followed by its efficiency analysis. Note that since it is only leaders that are actually stored in the Spatial Index Table, in the interest of simplicity we refer to objects and leaders interchangeably in the following presentation.

Algorithm 2 maintains two priority queues: $Q_{cell}$ for the spatial cells around $loc$, called cell candidates, and $Q_{obj}$ for the objects retrieved from cells in $Q_{cell}$, called neighbor candidates. The cells in $Q_{cell}$ are organized in ascending order by their distance from $loc$ (i.e., the shortest distance between $loc$ and any point in the cell), whereas objects in $Q_{obj}$ are organized in descending order by their distance to $loc$. In the beginning, both queues are empty. The first step is to push the cell that contains the $loc$, $c$, to $Q_{cell}$. Then, the two queues are to be iteratively used in such a way that, in each iteration we pop one cell $c$ in $Q_{cell}$ (i.e., the one closest to $loc$), then push those four cells that share an edge with $c$ to $Q_{cell}$, and add all objects in $c$ to $Q_{obj}$. The iteration terminates on the condition (line 2) that the closest cell in $Q_{cell}$ is farther than the first $k$ objects in $Q_{obj}$. Figure 8 shows an example after three iterations, where gray cells refer to those in $Q_{cell}$ (priority numbers are also given) and the three yellow cells are those whose objects are added into $Q_{obj}$ (the order of visiting is given by $a,b,c$). We use arrows to denote the distance from a cell to $loc$. The rationale behind our algorithm is that the distance between a cell and $loc$ is the lower boundary of the distance from any object within that cell to $loc$. Because of this, we use cell candidates to estimate the the distance of objects (in those cells) to $loc$ before actually retrieving the objects we wish to measure from the Spatial Index Table.

The efficiency of Algorithm 2 is mainly affected by the time required for recovering a certain number of nearest cells that contain enough moving objects. To improve the time for retrieving each cell, we utilize BigTable's feature of supporting very fast retrieval on a range of rows. We hope to retrieve all needed moving objects within a cell that is stored as a contiguous range of rows in the Spatial Index Table. Thus, we define a cell to be a $2^d \times 2^d$ space of spatial indexes in a nearest neighbor search, called a *NN cell*. That is, if a spatial index uses Hilbert Curves at level $l_s$ (i.e., $2^{l_s} \times 2^{l_s}$ spatial cells), the cells in a nearest neighbor search are the grids of the Hilbert Curve at level $l_n = l_s - d$, and by the property of Hilbert Curves, spatial indexes within a lower level are still consecutive. Figure 8 illustrates this property by splitting an NN cell into four spatial cells that correspond to a contiguous key range in BigTable. The level employed by the Hilbert Curve $l_n$ (called the *NN level*) is

**Algorithm 2** Nearest Neighbor Search

**Require:** *loc*, find nearest neighbors around *loc*
**Require:** $k > 0$, at most $k$ neighbors can be returned
**Require:** $l_n > 0$, where $l_n$ is the level of the cell (i.e., the search region unit)
1: $Q_{cell} \leftarrow$ new priority queue // to pop the nearest cell to *loc*
2: $Q_{cell}$.push(cell that contains *loc* in level $l_n$))
3: $dist_{max} \leftarrow \infty$ // $dist_{max}$ is the maximum distance a returned neighbor may be from *loc*
4: $Q_{obj} \leftarrow$ new priority queue // to pop the furthest object from *loc*
5: **while** $Q_{cell}$ is not empty **do**
6: $\quad c \leftarrow Q_{cell}$.pop()
7: $\quad$ **if** distance from $c$ to $loc > dist_{max}$ **then**
8: $\quad\quad$ break // the predictive version will consider velocity and predict duration in this pruning
9: $\quad$ **end if**
10: $\quad$ **for** *object* in $c$'s row of the Spatial Index Table **do**
11: $\quad\quad Q_{obj}$.push(*object*)
12: $\quad\quad$ **if** $Q_{obj}$.size() $> k$ **then**
13: $\quad\quad\quad Q_{obj}$.pop()
14: $\quad\quad$ **end if**
15: $\quad\quad$ **if** $Q_{obj}$.size() $= k$ **then**
16: $\quad\quad\quad dist_{max} \leftarrow$ distance from $Q_{obj}$.top() to *loc*
17: $\quad\quad$ **end if**
18: $\quad$ **end for**
19: $\quad$ **for** *neighborCell* around $c$ **do**
20: $\quad\quad Q_{cell}$.push(*neighborCell*)
21: $\quad$ **end for**
22: **end while**
23: Return *result*

**Algorithm 3** Calculate Best Search Level

**Require:** *loc*, nearest neighbor query center is *loc*
**Require:** $\rho_0$, best density, i.e., number of objects in best search level cell
**Require:** $n$, the number of moving objects in the whole space
1: $l_n \leftarrow \frac{1}{2} \log \frac{n}{\sigma}$
2: $min_{l_n} \leftarrow -\infty$
3: $max_{l_n} \leftarrow \infty$
4: **loop**
5: $\quad c \leftarrow$ the cell in level $l_n$ that contains *loc*
6: $\quad m \leftarrow$ number of objects in $c$
7: $\quad \delta = \frac{1}{2} \log \frac{m}{\sigma}$
8: $\quad$ **if** $\delta > 0$ **then**
9: $\quad\quad min_{l_n} \leftarrow l_n$
10: $\quad$ **else if** $\delta < 0$ **then**
11: $\quad\quad max_{l_n} \leftarrow l_n$
12: $\quad$ **end if**
13: $\quad l'_n \leftarrow \delta + l_n$
14: $\quad$ **if** $l'_n \leq min_{l_n}$ or $l'_n \geq max_{l_n}$ **then**
15: $\quad\quad$ break
16: $\quad$ **end if**
17: $\quad l_n \leftarrow l'_n$
18: **end loop**
19: Return $l_n$

a tunable parameter in our algorithm, which means the same nearest neighbors query can be completed with different cell sizes ($l_n$) without making any modifications to the Spatial Index Table.

### 3.4.2 Adaption to Spatial Density

The NN level is critical to the performance of nearest neighbor search. In this section, we present a method called Fast Level Adaptive Grid (FLAG) for fast tuning of NN levels as well as a caching scheme to reduce the overhead of calculating these NN levels. It is required that the tuning imposes no effect on update or other queries, so access to BigTable storage (i.e., the Spatial Index, Location, and Affiliation Tables) should be avoided when we tune or cache NN levels. Basically, we need to find an NN level so that every visited NN cell has $\sigma$ objects, or as close to $\sigma$ objects as possible. The value of $\sigma$ is a parameter decided by how the Spatial Index Table is stored in BigTable and is independent to the tuning algorithm.

*FLAG Algorithm:* Assuming that $\sigma$ is known, we give the calculation of NN levels in Algorithm 3. Initially, we guess the NN level $l_n$ by assuming all objects are uniformly distributed over the whole map (line 3). After the first estimation of object distribution, we check the number of objects in cell $c$ of level $l_n$ that contains *loc*. If the estimation is too large (resp. too small), we increase $l_n$ by $\delta = \frac{1}{2} \log \frac{m}{\sigma}$ to decrease (resp. increase) the objects in cell $c$ (again assuming that objects are uniformly distributed in $c$). The tuning continues until $l_n$ cannot be further adjusted.

Algorithm 4 illustrates how to cache the old NN levels of different areas for future use. Each cached NN level $l_n$ is associated with a range in the spatial index [*left, right*], which makes the caching location-sensitive. For example, an $l_n$ in urban and rural areas will most likely be different as there are probably more objects in a cell of an urban area than that of a rural one. The NN level $l_n$ is also associated with a timestamp showing when it was calculated. This is of great importance especially for business centers, where people stay only during working hours but leave after work.

When a nearest neighbor search targeting *loc* arrives, we search in the cache for a range [*left, right*] that covers *loc*, i.e., $h(loc) \in$ [*left, right*] ($h$ is the spatial index function). If it has not been cached or the timestamp is too old, a new $l_n$ will be calculated for *loc* by tuning algorithm (Algorithm 3). Then, we add a new cache record ($l_n$, [*left, right*]) where *left, right* are respectively the smallest and largest spatial indexes in cell $c$ at level $l_n$ that contains *loc*.

**Algorithm 4** Best Search Level Cache

**Require:** *loc*, nearest neighbor query center is *loc*
1: $index \leftarrow h(loc)$
2: $r \leftarrow$ find cache record ($l_n$, [*left, right*]) such that $index \in$ [*left, right*]
3: **if** $r$ exists and $r.created\_time$ is not too old **then**
4: $\quad$ Return $r.l_n$
5: **end if**
6: $l_n \leftarrow$ calculate best $l_n$ for *loc*
7: $c \leftarrow$ the cell in level $l_n$ that contains *loc*
8: add cache record ($l_n$, [*left*($c$), *right*($c$)])
9: Return $l_n$

## 3.5 Aged Data Archiving

Aged data should be written onto disk so that a history of moving patterns can be later analyzed for useful information such as travel paths and points of interest. When retrieving data for answering history queries, non-sequential and non-consecutive IOs are common. A naive scheme for archiving historical data is to flush an updated object location onto the disk before a new update arrives. This scheme can incur a large number of disk IOs, where each IO suffers from a latency penalty (i.e., seek time and rotational delay). To reduce latency overhead, double-buffering (or ping-pong buffering) can be employed. While updates are taking place on one memory buffer, another memory buffer is flushed onto the disk. What we must ensure in this scheme is that the time it takes to flush aged data from one buffer onto the disk is less than the time it takes to fill the other buffer in memory. Formally, let us denote the time to flush a buffer to the disk as $T_d$ and the time to fill a buffer as $T_m$, the following constraint must be observed: $\min T_m \geq \max T_d$.

At first glance, this double-buffering scheme seems to be simple to implement: When the location of an object is updated, we write that update to one of the two history buffers. When the current buffer is full, we flush it and then swap buffers. A major advantage of this scheme is keeping BigTable small: each object stores only one piece of location data. However, this simple implementation is inadequate for two reasons. First, since disk IO is much slower than memory IO, the archiving scheme may require flushing aged data onto parallel disks. Second, historical data are often analyzed by objects or by locations. To avoid scanning all files in response to an object or location query, preserving data locality must be considered. Therefore, we must design buffer strategy to meet the requirements of effective data locality and efficient parallel IOs.

As we mentioned in Section 3.3, we keep a number of historical locations in memory for each object. The purpose of this in-memory caching is for the support of applications such as travel-path rendering, current location positioning (via algorithms such as Viterbi [26]), and future location prediction. Thus, for each object we have $m$ in-memory records.

### 3.6 Parallel Archiving Streams

We present a parallel ping-pong scheme (PPP), which is designed to deal with three problems as follows:

1. How should the buffer be partitioned to preserve access locality?

2. What should the buffer-page size $s_B$ be for each archiving stream?

3. What should be the desired number of disks $n_d$?

#### 3.6.1 Data Placement

To address the problem of data partitioning, we view location data as a two-dimensional matrix, with rows representing objects and columns their locations. Because the location updating frequency of different objects vary, a column is copied to an aged-buffer page only when it is full. A buffer page is flushed onto disk only when enough objects have filled the page. Suppose we have $n_d$ buffer pages for $n_d$ disks. We would like to 1) flush an object to a minimal number of disks, and 2) place nearby objects on the same buffer page as much as possible.

To meet the locality requirement, we transfer data of object $i$ ($1 \leq i \leq n_o$) to disk $hash_d(i, loc_{i,0})(1 \leq hash_d(x,y) \leq d)$. This will guarantee that any object's archived data are always located on the same disk. Besides this object-locality, an object's initial location $loc_{i,0}$ is also considered in generating $hash_d$. This initial location is a good estimate of an object's potential location at later points of time because moving objects are unlikely to move too far away from their initial position after only a short period of time. For example, only a small subset of people travel to more than three cities in a single week. Over shorter periods of time, even city dwellers who typically restrict themselves to only a handful of locations in a single day often use the same transit routes.

#### 3.6.2 Determine $s_B$ and $n_d$

Since MOIST employs a double-buffering scheme, we need to reserve $s_B = s_{rec} \times n_o$ for caching aged data. In other words, the primary and secondary buffer reserves a space of $s_B$ for data in memory. These two buffers swap roles whenever the aged-data buffer has been flushed. Given $n_d$ disks, to balance workload, we assign each disk $n_o/n_d$ objects via hashing. Each disk then deals with a buffer of size $s_B/n_d$.

We measure cost based on disk utilization. There are two tasks which must be analyzed in relation to disk utilization: when aged data are written to parallel disks, and when queries to aged data are conducted (a disk read is performed) for pattern analysis. A large $n_d$ or small $s_B/n_d$ decreases the write-side of disk utilization. This can be explained as follows: let a disk's rotational delay and seek time be $T_{rot}$ and $T_{seek}$, respectively, and data transfer rate be $R_{disk}$. Then on each disk we have

$$T_d = T_{rot} + T_{seek} + s_B/(n_d \times R_{disk}). \quad (1)$$

The larger the value of $n_d$, the more the seek time and rotational delay dominate $T_d$. In this case, disk utilization $U_d$ can be expressed as $U_d = s_B/(n_d \times R_{disk} \times (T_{rot} + T_{seek}))$.

At the other hand, a small per-disk buffer size improves query-side locality, and hence increases the query-side (i.e., the read-side) of disk utilization. This can be explained by revisiting Eq.1. The larger the value of $n_d$, the better IO resolution from reading historical data for a set of objects is. In the worst case, when $n_d = 1$, finding historical data of an object requires reading the entire history archive, which can be several thousand times larger than that of $s_B$, spanning several days of history. Though some optimization can be performed, IO resolution remains deficient when the locality of writes is poor. In this case, the read resolution (i.e., read effectiveness) $R_d$ can be approximately expressed as

$$R_d = k \times n_d/n_o.$$

At one extreme, when $n_d = n_o$, resolution is perfect because each object has its own disk to cache its data. On the other end of the spectrum, when $n_d = 1$, resolution is at its worst. We scale the fraction $n_d/n_o$ by a normalization factor $k$, which can be tuned based on the real operational cost of the computer clusters, the fraction of write versus read operations, and the desired tradeoff between $U_d$ and $R_d$.

The optimization problem can then be formulated as maximizing overall disk utilization, $\min(U_d, R_d)$, given the constraint $\min T_m \leq \max T_d$, or

$$\max \min(U_d, R_d)$$

$$\text{subject to } \min T_m \geq \max T_d.$$

Since $U_d$ monotonously decreases and $R_d$ monotonously increases as $n_d$ increases, the maximum of $\min(U_d, R_d)$ is achieved when $U_d = R_d$ and $n_d$ is unconstrained. If $n_d$ which satisfies $U_d = R_d$ also satisfies $\min T_m \geq \max T_d$, we take this $n_d$ as the best configuration. Otherwise, the optimal $n_d$ shall be the one that satisfies $\min T_m = \max T_d$.

## 4. EXPERIMENTS

This section presents our experiments for evaluating MOIST. We designed experiments to answer two key questions:

1. How effectively can MOIST eliminate update redundancy (e.g., update shedding ratio)?

2. What are the limitations of BigTable for supporting MOIST (e.g., QPS of update and nearest neighbor query that BigTable can support for MOIST)?

Note that we conducted two sets of experiments separately:

1. When we performed experiments about MOIST's school, we tuned error bound and the shedding ratio, to see the performance change.

2. When we performed experiments about BigTable, the error bound was set to be zero, i.e., we did these experiments under the worst case: if every object is a leader, how is the performance of our basic indexing and kNN algorithm.

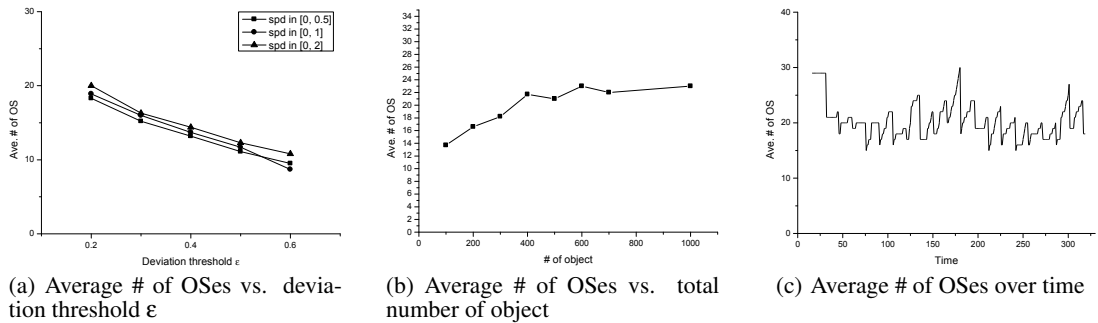

(a) Average # of OSes vs. deviation threshold ε

(b) Average # of OSes vs. total number of object

(c) Average # of OSes over time

**Figure 9: Impact of parameters on the average number of OS**

These two sets of experiments show that both our basic indexing and kNN algorithm and school have significant performance improvement over $B^x$-tree and other methods without school.

## 4.1 Datasets and Experiment Setup

We implemented and deployed MOIST in Google's data center. Each physical node was configured with a 1GHz CPU, 1GB RAM and a 1GB Disk. The MOIST server called Bigtable APIs for read and write operations. The overall BigTable resource quota was 2GB RAM, 300GB disk storage and 512KBps bandwidth. We used multiple machines to simulate traffic load as well as real services. All load tests were run by triggering updates and queries from multiple machines simultaneously: with up to 20,000 virtual machines, each running 50 threads, we were able to simulate a population of one million independent mobile clients. We regulated the update/query workload to MOIST by changing the update/query interval of each client. To simulate real servers, we deployed MOIST on multiple servers and studied the improvement on performance.

Due to the lack of real moving object datasets with a large population and high update frequency, we generated synthetic datasets of moving objects with positions on a square map of $1,000 \times 1,000$ units size. A similar simulation is used in many related works (e.g., [16]). To test MOIST's performance, we simulated real mobile object moving patterns in an urban area as much as possible. We used a road-networked map that had rectanglar buildings surrounded by roads. Each building was given an entrance. Moving objects were divided into two types: pedestrians and cars. We let each object initially move along a randomly selected road. Velocity was chosen between 0 and 1 units/second for pedestrians and between 1 and 2 units/second for cars. The locations and velocities in each update message were randomly perturbed to simulate noise, and the update interval was randomly chosen between zero and five seconds. When an object reached a crossroad, it chose a turn with equal probability. When a pedestrian was near an entrance to a building, they chose to enter it with 5% probability. Once inside a building, a pedestrian exited the building with a 5% probability also. During the time a pedestrian was inside of a building, each update would assign a position to the pedestrian within the building uniformly, at random.

We also studied the support of BigTable for location updates and nearest neighbor searches which showed that our scheme enjoyed a significant improvement, though it suffered from some limitations when compared to other memory- or disk-based methods. To stretch BigTable to its limit, updates and queries applied to a population of 400k to 1m objects with randomly chosen positions and velocities in a space size of 1km$^2$ were carried out. Since bottlenecks may take place in either BigTable or the server receiving the updates and queries, we also used multiple servers to receive updates and queries.

## 4.2 Effectiveness of MOIST

To test the effectiveness of MOIST, two competing aspects must be considered, the benefit of reductions in the number of updates versus the cost of clustering to obtain fewer OSes. To address this question, we considered the number of read and write operations performed by the server on BigTable (where the Location, Spatial Index and Affiliation Tables are stored), as this was the major bottleneck of our work. Generally, BigTable had a much better concurrency in read operations than write ones, so a reduction of *write operations* was of higher priority.

### 4.2.1 Update Reduction

Since only the leader of an object school (OS) incurs write operations to BigTable, we used the number of OSes as the benchmark for the benefits of update reduction. We only examined the average number of OSes for each clustering cell within a cluster interval, in order to avoid the consideration of the influence of reclustering (the influence of reclustering will be discussed in the Section 4.3).

In MOIST, three factors have the potential to impact the number of OSes: the deviation threshold ε, the average size of an OS, and the cluster interval $T_c$ for a cluster cell. As the the average size of an OS was not tunable, we increased the total number of objects, which is roughly the product of the average size of an OS and the average number of OSes. We separately investigated the influences of these factors and plot the results in Figure 9. We used a default update frequency of one update per second and a default population of 100 objects.

In Figure 9(a), the average number of OSes decreases linearly with deviation threshold ε. Meanwhile, the speed distribution has little impact on the number of OSes. Figure 9(b) indicates that there is little increase in the number of OSes when the number of objects increases by a factor of 10. Therefore, the update shed rate is able to achieve about 90% when 1,000 objects are in the space. Finally, Figure 9(c) shows that an update interval of $T_c = 10$ seconds can keep the variance of the number of OSes within 10.

### 4.2.2 Performance Tradeoffs of Clustering

We first investigate the latency of each clustering, a key metric of the clustering performance. The latency of one clustering in terms of the interaction with BigTable consists of three parts, read time, computation time and write time. Information needed by the clustering will be first retrieved in batch from the Spatial Index Table and Affiliation Table, which is in read time. Then, leaders are clustered into groups by the algorithm of Section 3.3.2, which is in

1846

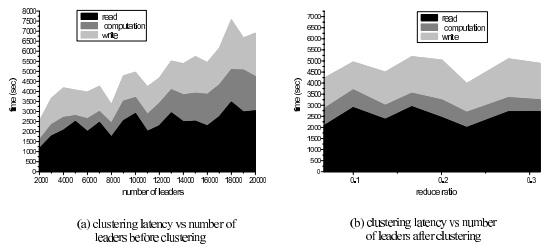

(a) clustering latency vs number of leaders before clustering

(b) clustering latency vs number of leaders after clustering

**Figure 10: Performance of clustering: per-clustering latency**

computation time. Finally, the Affiliation Table and the Spatial Index Table are updated according to the computation result, which is in write time. Figure 10(a) and Figure 10(b) show clustering latencies versus increasing numbers of pre-clustering leaders with fixed numbers of post-clustering leaders (i.e., 1k leaders), and fixed numbers of pre-clustering leaders (i.e., 10k leaders) with increasing number of post-clustering leaders, respectively. The plots also give the ratio of read, computation and write time for each setting. Figure 10(a) shows that the rising of latency depends more on the read time rather than the increasing number of leaders. Figure 10(b) shows that the latency has little to do with the reduction ratio of the settings in the plot.

We discuss the influences of clustering by learning the benefits of improving the nearest neighbor search QPS (NN QPS). More clusterings mean little objects (i.e., only leaders) in the Spatial Index Table, thus increasing NN QPS remarkably (e.g., NN QPS is doubled when the population in 1km$^2$ decreases from 10k to 1k). However, higher frequency of clustering will lead to more time cost, which will affect NN query and hence decrease the overall QPS. Here, we show our results about choosing a good clustering frequency. We assume two settings, setting A and B. For both settings, there were initially 1k leaders out of 20k objects. With each object updating their locations, we assume in setting A that the number of leaders increases linearly from 1k to 20k in 30sec, and in setting B that the number of leaders increases linearly from 1k to 20k in 60sec. Figure 11 plots the NN QPS variances of the two settings. Both settings A and B have roughly optimal frequencies of clustering that best improve the NN QPS (baseline of using no clustering is given by the black horizontal line). We use settings A and B to represent the cases where objects are moving in highly dynamic (leaders of A increase much faster than those in B) and in relatively fixed manner. Figure 11 shows that the optimal clustering frequency of setting A is a bit larger than that of B while the effect of clustering in setting A is better. Compared to the baseline of no clustering, the benefits of clustering in both settings are remarkable. While according to [6], the disk-based $B^x$ tree implementation needs a CPU time of at least 0.005-0.01 second, the peak NN QPS of our implementation has at least a 4x speedup than that of [6] in the same scenario.

### 4.3 Effectiveness of BigTable

In this section, we present the overall performance of our server that employs the BigTable infrastructure. We use the metric of Queries Per Second (QPS) to evaluate the performance.

#### 4.3.1 Adaptation over BigTable

BigTable has accessing criteria that is in stark contrast to page-based disk methods due to its distributed key-value storage scheme. In Section 3.4, we give an adaptive method called Fast Level Adaptive Grid (FLAG) for fast nearest neighbor search by considering object density skew when determining the granularity of search range.

We conducted our experiment on a map with no moving objects in order to evaluate the performance the nearest neighbor search aspect of MOIST over the search range. As discussed previously, a naive (fixed NN level) NN search algorithm will experience a quadratic rise in time cost as the search range limit increases. This was confirmed by figure 12(b), and QPS of fixed NN level drops drastically as figure 12(a) illustrates. From this, we can ascertain that an increase in performance can be obtained by a decrease of the NN level. Most importantly, this experiment proves that our Fast Level Adaptive Grid (FLAG) scheme works well in the presence of increasing search range. By adapting NN level automatically, FLAG maintains high performance even as range limit increases.

To evaluate robustness against an increase in density, we conducted the following experiment. From sparse to dense, 1K, 10K, 50K, 100K objects were uniformly and randomly placed into a square area of 1km$^2$ size. Figure 12(d) illustrates that a fixed NN level will experience a linear increase in time cost with the increase of density. As Figure 12(c) shows, QPS of fixed NN level drops when density increases. From these two figures, we can see that faster process time can be achieved by increasing NN levels when density increases. For FLAG, this experiment testifies that relatively high performance is conserved in face of increasing density.

#### 4.3.2 Single-Server QPS

In this experiment, we set up 10 machines that ran client simulators to query a single server. Each client simulator had 100 threads which concurrently queried the server. Before the test began, we configured the population and update frequency of each of the clients (i.e., their threads). Then, for each query generated by a thread, a random object id ($1 \leq id \leq$ # of objects) would be assigned to the object.

Figure 13(a) reveals that update QPS decreases very little as the number of indexed objects increase. Even in the case of a single front-end server and 1 million indexed objects, our design can process as many as 7,875 update requests per second. The amortized time cost for a single update request is less than 0.2ms, which surpasses state-of-the-art disk-based designs such as the $B^x$ trees evaluated in [6].

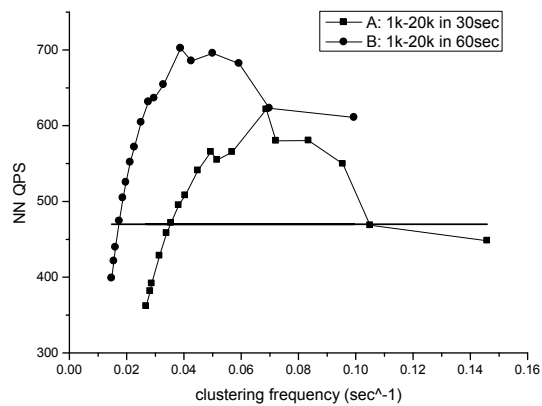

**Figure 11: Influence of clustering: improvement of nearest neighbor search QPS (NN QPS). Black horizontal line gives the NN QPS with no clustering.**

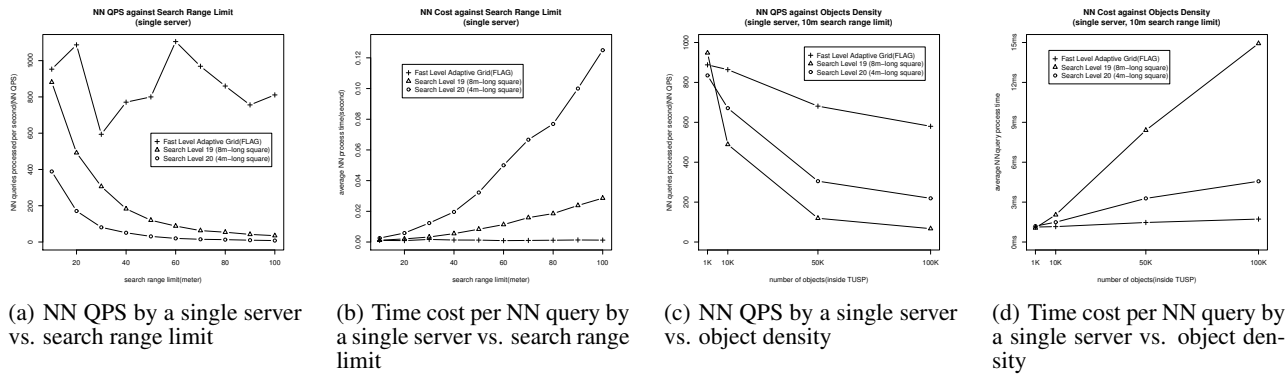

(a) NN QPS by a single server vs. search range limit

(b) Time cost per NN query by a single server vs. search range limit

(c) NN QPS by a single server vs. object density

(d) Time cost per NN query by a single server vs. object density

Figure 12: Effectiveness of adaptation over BigTable using FLAG

### 4.3.3 Multiple-Server QPS

Theoretically, MOIST has very little communication overhead with the increase in the number of machines. Furthermore, even with restricted BigTable resources, the performance of MOIST scales in a favorable fashion with the number of machines.

Figure 13(b) exhibits the QPS when we deployed 5 servers sharing a single BigTable to handle updates/queries. The speedup is very close to the optimal case, i.e., a 5x speedup. Figure 13(c) shows the QPS when we deployed 10 servers sharing a single BigTable to handle updates/queries. Although the QPS is not very stable over time, the average and maximum speedups are very close to optimal. In spite of this, the QPS increases to 60k with the deployment of 10 machines, which accounts for almost an 80x speedup over $B^x$ trees [6].

## 5. APPLICATIONS

In this section, we give a brief introduction on the applications that we have deployed or plan to deploy based on the MOIST. We currently focus on two main fields, realtime transit and realtime coupon.

Realtime transit targets to organize the realtime locations of public transportation vehicles, like buses and taxis, and makes them easily accessible by users. Taking a bus as an example, most urban residents take it every day, however, they are frustrated by its unpredictable arrival time due to bad traffic. If they can search the bus location on their mobile phone, they probably can save a lot of time in waiting at bus stops. From the aspect of implementation, buses' GPS locations should be updated to server, and the server takes charge of storing location data, building an index and responding to users' location queries. Since buses can travel as far as hundreds of meters in one minute, the location updating frequency should be high enough. This is just one of problems that MOIST can solve, additionally, its index can support users to browse all running buses near a location. We've built a Bus Alert Service, and a corresponding Android application through which users can use the service. This work was first prototyped in Taipei in March 2011. The server serves about 5000 buses' realtime locations, and each bus updated its GPS location twice a minute. From the client, users could (1) query a bus' location, (2) browse all buses nearby, and (3) set an alarm to remind if the selected bus is approaching. In October 2011, we launched the realtime location service of Google shuttles as a key feature of Google Campus, which is widely used inside Google to access company resources.

Another field where we want to apply MOIST is realtime coupon. On the one hand, users are encouraged to update their locations to get coupons from nearby shops or restaurants. On the other hand, shop owners can submit coupons to our system, targeting nearby users immediately. Proposing an interesting scenario: a user is wandering then lunch time comes, he gets out his phone, registers for coupons of nearby Chinese restaurants. At the same time, a pretty good Chinese restaurant within 300 meters finds that there are several seats open, and submits several of 20%-off coupons targeting customers within 1,000 meters. Our service will match the restaurant and customer accurately. In all, this is an interesting idea, and we'd like to deploy it in the near future.

MOIST can be employed by many kinds of applications as long as they need frequent location updates. Furthermore, it can be used to discover points of interests or even automatically draw maps based on mining the large scale user location updates for a long time. We've deployed MOIST as one of the fundamental location-based infrastructures, and we can imagine all kinds of exciting applications coming afterwards.

## 6. CONCLUSION

In this paper, we presented MOIST, a recursive spatial partitioning indexer built upon BigTable, which significantly improves performance for spatial NN and history queries. We showed that through object-schooling, update frequency can be significantly reduced, and through data placement and our proposed parallel ping-pong scheme, we can archive historical data for efficient data processing. Through our experiments, we have shown that MOIST is able to scale well with increasing, large volume location updates with low latency. We have also developed a *Transit Alert* prototype on top of MOIST for keeping track of real-time locations of buses and users. Our future work will focus on utilizing historical data to conduct pattern analysis for improving user experience on LBS applications such as route planning, map makers, and point-of-interest data mining.

Besides MOIST, the team [1] has been developing and prototyping indoor positioning and navigation technologies using WIFI signals, inertial navigation systems (INS) [10], and imagery data [4]. Schemes of INS calibration, signal processing, signal fusion, and crowd sourcing will be deployed together with MOIST to provide robust indoor/outdoor location-based services.

## 7. REFERENCES

[1] http://sitescontent.google.com/mobile-2014-research.
[2] L. Biveinis, S. Šaltenis, and C. Jensen. Main-memory operation buffering for efficient R-tree update. In *VLDB*, pages 591–602, 2007.

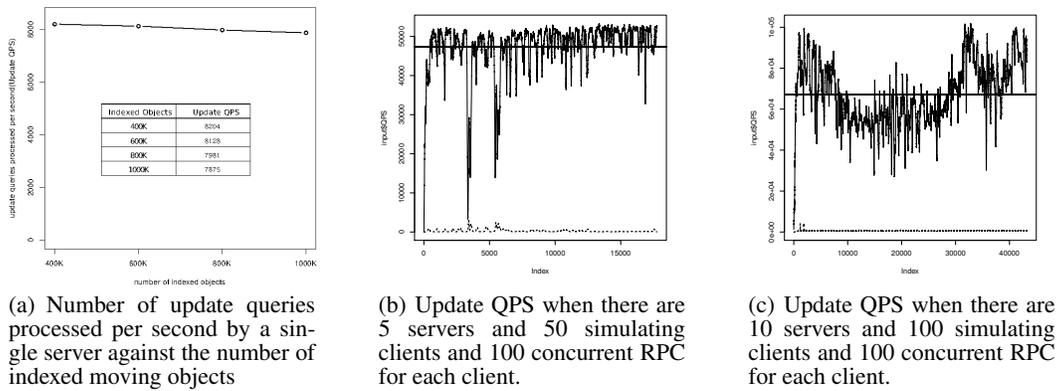

(a) Number of update queries processed per second by a single server against the number of indexed moving objects

(b) Update QPS when there are 5 servers and 50 simulating clients and 100 concurrent RPC for each client.

(c) Update QPS when there are 10 servers and 100 simulating clients and 100 concurrent RPC for each client.

Figure 13: Impact of parameters on the average number of OS. For (b) and (c), $x-$axis: timeline; dashed line: the number of failed queries per second (these queries are not included in QPS); horizontal line: the overall average QPS

[3] Y. Cai, K. Hua, G. Cao, and T. Xu. Real-time processing of range-monitoring queries in heterogeneous mobile databases. *IEEE Transactions on Mobile Computing*, 5(7):931–942, 2006.

[4] E. Y. Chang. *Foundations of Large-Scale Multimedia Information Management and Retrieval: Mathematics of Perception*. Springer, 2011.

[5] F. Chang, J. Dean, S. Ghemawat, W. Hsieh, D. Wallach, M. Burrows, T. Chandra, A. Fikes, and R. Gruber. Bigtable: A distributed storage system for structured data. *ACM Transactions on Computer Systems (TOCS)*, 26(2):1–26, 2008.

[6] S. Chen, C. Jensen, and D. Lin. A benchmark for evaluating moving object indexes. *PVLDB*, 1(2):1574–1585, 2008.

[7] S. Chen, B. Ooi, K. Tan, and M. Nascimento. ST2B-tree: a self-tunable spatio-temporal b+-tree index for moving objects. In *ACM SIGMOD*, pages 29–42, 2008.

[8] S. Chen, B. Ooi, and Z. Zhang. An adaptive updating protocol for reducing moving object database workload. *PVLDB*, 3(1-2):735–746, 2010.

[9] A. Civilis, C. Jensen, and S. Pakalnis. Techniques for efficient road-network-based tracking of moving objects. *IEEE Transactions on Knowledge and Data Engineering*, 17(5):698–712, 2005.

[10] Y. Gao, Q. Yang, G. Li, E. Chang, D. Wang, C. Wang, H. Qu, P. Dong, and F. Zhang. Xins: the anatomy of an indoor positioning and navigation architecture. In *Proceedings of the 1st International Workshop on Mobile Location-based Service*, pages 41–50, 2011.

[11] B. Gedik, K. Wu, P. Yu, and L. Liu. Mobiqual: Qos-aware load shedding in mobile cq systems. In *ICDE*, pages 1121–1130, 2008.

[12] S. Har-Peled. Clustering motion. *Discrete and Computational Geometry*, 31(4):545–565, 2004.

[13] H. Hu, J. Xu, and D. Lee. A generic framework for monitoring continuous spatial queries over moving objects. In *ACM SIGMOD*, pages 479–490, 2005.

[14] A. Jain, E. Chang, and Y. Wang. Adaptive stream resource management using kalman filters. In *ACM SIGMOD*, pages 11–22, 2004.

[15] C. Jensen, D. Lin, and B. Ooi. Query and update efficient B+-tree based indexing of moving objects. In *VLDB*, pages 768–779, 2004.

[16] C. Jensen, D. Lin, and B. Ooi. Continuous clustering of moving objects. *IEEE Transactions on Knowledge and Data Engineering*, 19(9):1161–1174, 2007.

[17] M. Kurashige and S. Fukushima. Image converter for mapping a two-dimensional image onto a three dimensional curved surface created from two-dimensional image data, Nov. 8 1994. US Patent 5,363,476.

[18] Y. Li, J. Han, and J. Yang. Clustering Moving objects. In *KDD*, pages 617–622, 2004.

[19] D. Lin, C. Jensen, B. Ooi, and S. Šaltenis. Efficient indexing of the historical, present, and future positions of moving objects. In *Proceedings of the 6th International Conference on Mobile Data Management*, pages 59–66, 2005.

[20] M. Mokbel and W. Aref. Space-Filling Curves. 2008.

[21] P. Pesti, L. Liu, B. Bamba, A. Iyengar, and M. Weber. RoadTrack: scaling location updates for mobile clients on road networks with query awareness. *PVLDB*, 3(1-2):1493–1504, 2010.

[22] S. Prabhakar, Y. Xia, D. Kalashnikov, W. Aref, and S. Hambrusch. Query indexing and velocity constrained indexing: Scalable techniques for continuous queries on moving objects. *IEEE Transactions on Computers*, 51(10):1124–1140, 2002.

[23] S. Šaltenis, C. Jensen, S. Leutenegger, and M. Lopez. Indexing the positions of continuously moving objects. In *ACM SIGMOD*, pages 331–342, 2000.

[24] Y. Tao, D. Papadias, and J. Sun. The TPR*-tree: An optimized spatio-temporal access method for predictive queries. In *VLDB*, pages 790–801, 2003.

[25] K. Tzoumas, M. Yiu, and C. Jensen. Workload-aware indexing of continuously moving objects. *PVLDB*, 2(1):1186–1197, 2009.

[26] A. Viterbi. A personal history of the Viterbi algorithm. *IEEE Signal Processing Magazine*, 23(4):120–142, 2006.

[27] M. Yiu, Y. Tao, and N. Mamoulis. The B dual-Tree: indexing moving objects by space filling curves in the dual space. *VLDB Journal*, 17(3):379–400, 2008.

[28] M. Zhang, S. Chen, C. Jensen, B. Ooi, and Z. Zhang. Effectively indexing uncertain moving objects for predictive queries. *PVLDB*, 2(1):1198–1209, 2009.